\documentclass[12pt]{article}
\usepackage{curves}
\usepackage{rotating}
\newcommand{\dd}{\mbox{\rm d}}

\newcommand{\oo}{\over}
\newcommand{\p}{\partial}

\newcommand{\be}{\begin{equation}}
\newcommand{\ee}{\end{equation}}
\newcommand{\lbl}{\label}
\newcommand{\bi}{\bibitem}
\newcommand{\ci}{\cite}

\newcommand{\vs}{\vspace}
\newcommand{\hs}{\hspace}

\def\bear{\begin{eqnarray}}
\def\ear{\end{eqnarray}}

\begin{document}
\baselineskip .7cm

\thispagestyle{empty}

\ 

\vs{1.5cm}

\begin{center}

{\bf \LARGE Clifford Space as the Arena for Physics}

\vs{6mm}

Matej Pav\v si\v c\footnotetext{Email: MATEJ.PAVSIC@IJS.SI}

Jo\v zef Stefan Institute, Jamova 39, SI-1000 Ljubljana, Slovenia

\vs{1.5cm}

ABSTRACT
\end{center}

A new theory is considered according to which extended objects in
$n$-dimensional
space are described in terms of multivector coordinates which are interpreted 
as generalizing the concept of centre
of mass coordinates. While the usual centre of mass is a point,
by generalizing the latter concept, we associate with every extended object a 
set of 
$r$-loops, $r=0,1,..., n-1$, enclosing oriented $(r+1)$-dimensional surfaces
represented by Clifford numbers called $(r+1)$-vectors or multivectors. 
Superpositions of multivectors
are called polyvectors or Clifford aggregates and they are elements of
Clifford algebra. The set of all possible polyvectors forms a manifold,
called $C$-space. We assume that the arena in which physics takes
place is in fact not Minkowski space, but $C$-space. This has many
far reaching physical implications, some of which are discussed
in this paper. The most notable is the finding that although
we start from the constrained relativity in $C$-space we arrive at
the unconstrained Stueckelberg relativistic dynamics in Minkowski space
which is a subspace of $C$-space.

\vs{1cm}


\newpage

\section{Introduction}

A space (in particular, spacetime) consists of points (events).
But besides points there are also lines, surfaces, volumes, etc..
Description of such geometric objects has turned out to be very
elegant if one employs multivectors which are the outer
products of vectors \ci{Hestenes}. Multivectors and their linear combinations,
polyvectors, are elements of Clifford
algebras. Since in physics we do not consider point particles only, 
but also extended objects, it appears natural to consider
Clifford algebra as an arena in which physics takes place 
\ci{Pezzaglia,Castro,Pavsic1}.
Instead of spacetime we thus consider a more general space,
the so called {\it Clifford manifold} or {\it $C$-space}.
This is a space of the oriented $(r+1)$-dimensional areas enclosed 
by $r$-loops. In this paper I will show that extended objects can be described
by $r$-loops and that the $r$-loop coordinates are natural generalizations
of the concept of the center of mass coordinates. Besides the
center of mass velocity an extended object has also the area velocity,
the volume velocity, etc. (called multivector or holographic velocities). 
We generalize the theory of relativity from Minkowski space $M_4$ to 
$C$-space and thus bring into the game the holographic velocities.
Besides the speed of light a  fundamental length $L$ has to 
be introduced. If we take $L$ equal to the Planck length we find
that the maximum holographic speeds are very slow and this explains why on 
the macroscopic scale we do not observe them. For instance, the
volume (the 3-vector) speed is of the order of $10^{-62} m^3/sec$.

The action for a ``point particle'' in $C$-space is analogous to the
action for a point particle in Minkowski space $M_4$. It is equal to the
length of the world line in $C$-space. This action constrains
the polymomentum to the mass shell in $C$-space. If we reduce
the $C$-space action with respect to the 4-volume (4-vector or
pseudoscalar) variable $s=X^{0123}$, then all other variables
are independent and evolve with respect to $s$ which assumes the
role of evolution parameter (the true time). The action so reduced
is equivalent to the well known Stueckelberg action of the
relativistic dynamics \ci{Stueckelberg}. 

In the unconstrained theory, minimal length, action, the variables
$X^A = ({\sigma}, X^{\mu},X^{\mu \nu}, X^{\mu \nu \alpha},
X^{\mu \nu \alpha \beta})$ are functions of an arbitrary parameter $\tau$.
The 4-volume also changes with $\tau$. This explains why the world lines
(actually the world tubes, if particles are extended) in $M_4$
are so long along time-like directions, and have so narrow 
space-like extension. This is just a very natural ``final" state 
of objects evolving in $M_4$. Initially the objects may have
arbitrary shapes, but if their 2-vector and 3-vector speeds are
of the right sign (so it is on average in half of the cases), then 
their extensions along time-like directions will necessarily increase
for positive 4-vector speeds; increasing 4-volume necessarily implies
increasing length of a world tube (whose effective 2-area and 3-volume
are constant or decreasing). Long world lines are necessary in
order to provide the observed electromagnetic fields. Finite
extensions of world lines provide a cutoff to the electromagnetic
interaction which is predicted to change with time.

All the conservation laws are still valid, since the true physics
is now in $C$-space. In $C$-space there is no ``flow of time'' at all:
past, present and future coexist in the 16-dimensional ``block'' 
$C$-space, with objects corresponding to wordlines in $C$-space.
But, if the 4-dimensional Minkowski space $M_4$ is considered
as a slice which moves through $C$-space, objects on $M_4$ 
are evolving with respect to the Lorentz invariant evolution
parameter $s$. There is then a genuine dynamics on $M_4$ which is
induced by the postulated motion of $M_4$ through $C$-space. A reconsiliation
between two seemingly antagonistic
views is achieved, namely between the assertion that there is no
time, no flow of time, etc., and the view that there is evolution,
passage of time, relativistic dynamics. Both groups of thinkers are right, 
but each in its own space, and the link between the two views is in
the postulated motion of $M_4$. So far an important argument against relativistic 
dynamics has been that it implies the entire spacetime moving in a 5-dimensional
space and thus it introduces a ``meta time". In this respect relativistic
dynamics has been considered no better than the usual relativity together with
the assumption that a ``time slice" moves through $M_4$. Why to introduce a
5-dimensional space\footnote{We are here not talking about the extra dimensions
required by Kaluza-Klein theories or string theories, but about an extra
dimensions introduced in order to describe evolution.}, and why should the game
stop at five and not at six or more dimensions? Why not just stay with
4-dimensions of Minkowski space? Clifford space resolves those puzzles, since
Clifford space is defined over 4-dimensional spacetime, it does not
introduce extra dimensions of spacetime.

\section{Geometry of spacetime and Clifford algebra}

Usually it is  assumed that physics takes place in the ordinary spacetime
$V_n$, a manifold whose points are assigned the coordinates $x^\mu$,
$\mu = 0,1,2,...,n-1$. Generally a spacetime or any manifold can be
elegantly described by a set of basis vectors $\gamma_\mu$ satisfying
the Clifford algebra relations
\be
   \gamma_\mu \cdot \gamma_\nu \equiv \mbox{${1\oo 2}$} (\gamma_\mu
   \gamma_\nu + \gamma_\nu \gamma _\mu ) = g_{\mu \nu}
\lbl{1}
\ee
where $g_{\mu \nu}$ is the metric. In general $\gamma_\mu$ and thus
$g_{\mu \nu}$ can depend on $x^\mu$ (see e.g., refs. \ci{Hestenes,Pavsic1}).
In terms of $\gamma_\mu$ a vector $a$ in $V_n$ can be expressed as
\be
          a = a^\mu \gamma_\mu
\lbl{2}
\ee
Its square is then 
\be
    a^2 = a^\mu a^\nu \gamma_\mu \gamma_\nu = a^\mu a^\nu g_{\mu \nu}
    \equiv a^\mu a_\mu
\lbl{3}
\ee

A vector can in general be a field $a = a(x)= a^\mu (x) \gamma_\mu$
which depends on position. In particular, we may consider such a field
$a(x)$ whose components are position coordinates:
\be
          a(x) = x = x^\mu \gamma_\mu
\lbl{4}
\ee

Acting on $x$ by the differential operator d which acts on components, 
but leave the basis vectors unchanged (see \ci{Hestenes,Pavsic1}) we obtain
\be
  {\rm d} x = \dd x^\mu \gamma_\mu
\lbl{5}
\ee
This is an example of a {\it vector} in spacetime. Another example
is the velocity $v = (\dd x^\mu/\dd \tau) \gamma_\mu$.

The well known important quantity is the square
\be
    \dd x^2 = (\dd x^\mu \gamma_\mu)^2 = g_{\mu \nu} \dd x^\mu \dd x^\nu
\lbl{6}
\ee
which is the infinitesimal distance between two infinitesimally separated
points (``events") in $V_n$.

Another quantity is the square of the coordinate vector field
\be
   a^2(x) = x^2 = (x^\mu \gamma_\mu)^2 = g_{\mu \nu} x^\mu x^\nu 
\lbl{7}
\ee
which has a significant role in special relativity which acts in {\it flat}
spacetime with the  Minkowski metric $g_{\mu \nu} = \eta_{\mu \nu}$.

In a manifold we do not have points only. There are also lines,
2-dimensional surfaces, 3-dimensional surfaces, etc.. Description of all
those objects requires use of higher degree vectors, i.e., {\it
multivectors}, such as
     $$\gamma_\mu \wedge \gamma_\nu \equiv \mbox{${1\oo {2!}}$}
      (\gamma_\mu \gamma_\nu - \gamma_\nu \gamma_\mu) \equiv
     \mbox{${1\oo {2!}}$} [\gamma_\mu , \gamma_\nu]$$
\be
   \gamma_\mu \wedge \gamma_\nu \wedge \gamma_\alpha \equiv \mbox{${1\oo {3!}}$}
   [\gamma_\mu , \gamma_\nu , \gamma_\alpha]
\lbl{8}
\ee
\be       {\rm etc.}
\ee

An {\it oriented line element} is given by a vector $\dd x$ (eq.(\ref{5})).

An {\it oriented area element} is given by a bivector which is the wedge
product of two different vectors $\dd x_1$ and $\dd x_2$:
\be
     \dd x_1 \wedge \dd x_2 = \dd x_1^\mu \dd x_2^\nu \, \gamma_\mu \wedge
     \gamma_\nu
\lbl{9}
\ee
Similarly for an arbitrary multivector
\be
     \dd x_1 \wedge \dd x_2 \wedge ... \wedge \dd x_r =
     \dd x_1^{\mu_1} ... \dd x_r^{\mu_r} \, \gamma_{\mu_1} \wedge ...
     \wedge \gamma_{\mu_r}
\lbl{10}
\ee
If $r$ is less then the dimension of the space $V_n$, then the above 
multivector represents an $r$-dimensional {\it area element}. If $r=n$,
then (\ref{10}) becomes a {\it volume element} in the space $V_n$
\be
     \dd x_1 \wedge ... \wedge \dd x_n = \dd x_1^{\mu_1} ... \dd x_n^{\mu_n}
     \gamma
\lbl{11}
\ee
where the $n$-vector
\be
    \gamma \equiv \gamma_{\mu_1} \wedge ... \wedge \gamma_{\mu_n}
\lbl{12}
\ee
is called {\it pseudoscalar}. It is the highest possible $r$-vector in a space
of dimension $n$, since and $(n+1)$-vector is identically zero.

\section{Multivectors associated with finite $r$-surfaces}

Let us now consider $\dd x$ of eq.(\ref{5}) as the tangent vector to a curve
$X^\mu (\tau)$. Integrating over the curve we have after a suitable choice
of the integration constant that
\be
    \int \dd x = \int \dd \tau {{\dd X^\mu}\oo {\dd \tau}} \gamma_\mu
    = x^\mu \gamma_\mu
\lbl{13}
\ee
where we have considered a {\it flat spacetime} and such a coordinate
system in which $\gamma_\mu$ are constants. The result of integration
in eq.(\ref{13}) is independent of the chosen curve $X^\mu (\tau)$. It
depends only on the initial and final point. In (\ref{13}) we have
taken the initial point at the coordinate origin and assigned to the
final point coordinates $x^\mu$.

Similarly we can consider $\dd x_1 \wedge \dd x_2$ as a tangent bivector to
a surface $X^\mu (\sigma^1,\sigma^2)$. After the integration over a chosen range
of the parameters $\sigma^1,~\sigma^2$ we obtain
\bear
    \int \dd x_1 \wedge \dd x_2 &=& \int \dd \sigma^1 \dd \sigma^2 \, {{\p X^\mu}
    \oo {\p \sigma^1}} {{\p X^\nu} \oo {\p \sigma^2}} \gamma_\mu \wedge \gamma_\nu
    \nonumber \\
     &=& {1\oo2 } \int \dd \sigma^1 \dd \sigma^2 \, \left ({{\p X^\mu}
    \oo {\p \sigma^1}} {{\p X^\nu} \oo {\p \sigma^2}} - {{\p X^\nu}
    \oo {\p \sigma^1}} {{\p X^\mu} \oo {\p \sigma^2}} \right )
    \gamma_\mu \wedge \gamma_\nu \nonumber \\
    &=& \mbox{${1\oo 2}$} X^{\mu \nu} \gamma_\mu \wedge \gamma_\nu 
\lbl{14}
\ear
where
\be 
    X^{\mu \nu} =  \int \dd \sigma^1 \dd \sigma^2 \, \left ({{\p X^\mu}
    \oo {\p \sigma^1}} {{\p X^\nu} \oo {\p \sigma^2}} - {{\p X^\nu}
    \oo {\p \sigma^1}} {{\p X^\mu} \oo {\p \sigma^2}} \right )  
\lbl{15}
\ee
By the Stokes theorem this is equal to
\be
    X^{\mu \nu} = {1\oo 2} \oint \dd \zeta \, \left ( X^\mu {{\p X^\nu}\oo
    {\p \zeta}} - X^\nu {{\p X^\mu}\oo {\p \zeta}} \right )
\lbl{16a}
\ee
where the integral runs along a loop given by the embedding functions
$X^\mu (\zeta)$ of a single parameter $\zeta$.

The inegral (\ref{14}) is a finite bivector and it does not depend on
choice of the surface enclosed by the loop. There is a family of loops
which all give the same result $X^{\mu \nu}$. Within the family there exists
a subset of loops with all their points situated on a plane. Amongst them
we may choose a representative loop.
We then interprete $X^{\mu \nu} $ as bivector coordinates of a representative
loop enclosing a flat surface element.

Choosing a coordinate system on the surface such that $\sigma^1 =X^1,~\sigma^2
= X^2$ we find ---when our surface is a flat rectangle--- that
     $$X^{12} = \int \dd X^1 \, \dd X^2 = X^1 \, X^2$$
\be     
     X^{21} = - \int \dd X^1 \, \dd X^2 = - X^1 \, X^2
\lbl{16}
\ee
     
Similarly, if we choose $\sigma^1,~ \sigma^2$ equal to some other pair
of coordinates
$X^\mu, ~X^\nu$, we obtain other components of $X^{\mu \nu}$ expressed
as the products of $X^\mu$ and $X^\nu$. In 4-dimensional spacetime we have
\be
      X^{\mu \nu} = \pmatrix{  0  & X^0 X^1  &  X^0 X^2  &  X^0 X^3 \cr
                             -X^1 X^0 & 0 & X^1 X^2 & X^1 X^3 \cr
                             -X^2 X^0 & -X^2 X^1 & 0 & X^2 X^3  \cr
                             -X^3 X^0 & -X^3 X^1 & - X^3 X^2 & 0 \cr }
\lbl{17}
\ee
These are just the holographic projections of our oriented surfaces onto the
coordinate planes.

In analogous way we can calculate  higher multivector components
$X^{\mu_1 \mu_2 ... \mu_r}$ for arbitrary $r \le n$.

A {\it vector} $x\equiv X_1 = X^\mu \gamma_\mu$ can be used to denote
a {\it point}, and $X^\mu$ are {\it coordinates} of the point. The 
{\it distance}
of the point from the coordinate origin is given by the square root of the
expression (\ref{7}). In other words, the {\it length} of line from the 
coordinate origin to the point is $(g_{\mu \nu} X^\mu X^\nu)^{1/2}$, and
the length square in flat spacetime is
\be
     |X_1|^2 \equiv X_1 * X_1 = (X^0)^2 - (X^1)^2 - (X^2)^2 - (X^3)^2
\lbl{17a}
\ee

Analogously, a bivector $X_2 = {1\oo 2} X^{\mu \nu} 
\gamma_\mu \wedge \gamma_\nu$
denotes a 1-{\it loop} enclosing a flat surface element, 
and $X^{\mu \nu}$ are (bivector type) coordinates
of the loop. The area that the loop encloses is given by the square root
of the scalar product between $X_2 = {1\oo 2} X^{\mu \nu} 
\gamma_\mu \wedge \gamma_\nu$
and its Hermitian conjugate or reverse $X_2^{\dagger} = {1\oo 2} X^{\mu \nu}
\gamma_\nu \wedge \gamma_\mu$:
\bear
    X_2^{\dagger} * X_2 &=& \mbox{${1\oo 4}$} [X^{\mu \nu} \gamma_\nu
   \wedge \gamma_\mu ] * [X^{\alpha \beta} \gamma_\alpha \wedge 
    \gamma_\beta] \nonumber \\
    &=& \mbox{${1\oo 4}$} X^{\mu \nu} X^{\alpha \beta} (\gamma_\nu
    \wedge \gamma_\mu)\cdot (\gamma_\alpha \wedge \gamma_\beta) \nonumber \\
    &=& \mbox{${1\oo 4}$} X^{\mu \nu} X^{\alpha \beta} \left [
    (\gamma_\mu \cdot \gamma_\alpha)(\gamma_\nu \cdot \gamma_\beta) -
    (\gamma_\mu \cdot \gamma_\beta)(\gamma_\nu \cdot \gamma_\alpha) \right ]
    \nonumber \\
    &=& \mbox{${1\oo 4}$} X^{\mu \nu} X^{\alpha \beta} (g_{\mu \alpha}
    g_{\nu \beta} - g_{\mu \beta} g_{\nu \alpha}) = \mbox{${1\oo 2}$}
    X^{\mu \nu} X_{\mu \nu}
\lbl{18}
\ear
In a four-dimensional Minkowski space with signature $(+ - - -)$ we have
\be
    |X_2|^2 = - (X^{01})^2 - (X^{02})^2 - (X^{03})^2 + (X^{12})^2 +
    (X^{13})^2 + (X^{23})^2
\lbl{19}
\ee
which is the Pythagora rule for surfaces, analogous to the one for lines.

An $r$-{\it vector} $X_r = {1\oo {r!}} X^{\mu_1...\mu_r} \gamma_{\mu_1}
\wedge \gamma_{\mu_2} \wedge ...
 \gamma_{\mu_r}$ denotes an oriented $r$-{\it surface} enclosed by
an $(r-1)$-{\it loop}, where $X^{\mu_1...\mu_r}$ are ($r$-vector type)
{\it coordinates} of the $r$-surface.

The precise shape of  the $(r-1)$-loop is not determined by the $r$-vector
coordinates $X^{\mu_1...\mu_r}$. Only the orientation and the {\it area}
enclosed by the loop are determined. These are thus {\it collective
coordinates} of a loop. They do not describe all the degrees of freedom
of a loop, but only its collective degree of freedom---area and orientation---
common to a family of loops.

Multivectors are elements of the {\it Clifford
algebra} generated by the set of basis vectors $\gamma_{\mu_1},...,
\gamma_{\mu_r}$. A generic {\it Clifford number} or {\it polyvector}
(also called Clifford aggregate) is a sum of multivectors:
\be
    X = \sigma + X^\mu \gamma_\mu + {1\oo {2!}} X^{\mu_1 \mu_2} \gamma_{\mu_1} 
    \wedge \gamma_{\mu_2} + ... + {1\oo {n!}} X^{\mu_1...\mu_n}
\gamma_{\mu_1} \wedge ... \wedge \gamma_{\mu_n}
\lbl{23}
\ee
where $X^A = (\sigma, X^\mu,X^{\mu},...)$ and $E_A = ({\underline 1},
\gamma_\mu, \gamma_{\mu \nu},...)$.

Taking $n=4$ and using the relations
\be
      \gamma_\mu \wedge \gamma_\nu \wedge \gamma_\rho \wedge \gamma_\sigma
      = I \epsilon_{\mu \nu \rho \sigma}
\lbl{23a}
\ee
\be
      \gamma_\mu \wedge \gamma_\nu \wedge \gamma_\rho = 
      I \epsilon_{\mu \nu \rho \sigma} \gamma^\sigma
\lbl{23b}
\ee
\be
      I \equiv \gamma = \gamma_0 \wedge \gamma_1 \wedge \gamma_2 \wedge
      \gamma_3
\lbl{23c}
\ee
where $\epsilon_{\mu \nu \rho \sigma}$ is the totally antisymmetric tensor,
and introducing the new coefficients
\be
    \xi_\sigma \equiv {1\oo {3!}}
    X^{\mu \nu \rho} \epsilon_{\mu \nu \rho \sigma} \; ,
    \qquad s \equiv {1\oo {4!}}
    X^{\mu \nu \rho \sigma} \epsilon_{\mu \nu \rho \sigma}
\lbl{23d}
\ee       
we can rewrite (\ref{23}) in terms of pseudovector and pseudoscalar
variables
\be
      X = \sigma + X^\mu \gamma_\mu + {1\oo 2} X^{\mu \nu} \gamma_\mu \wedge
      \gamma_\nu + \xi^\mu I \gamma_\mu + s I
\lbl{24}
\ee

Summation of multivectors of different degrees $r$ is no more curious than
the summation of real and imaginary numbers. The set of all possible polyvectors
$X$ forms a manifold, called {\it Clifford space}, or briefly, $C$-{\it space}.

\section{Multivectors associated with physical objects}

We will now show that
the $r$-vector coordinates $X^{\mu_1...\mu_r}$ when considered as describing
a physical object---not necessarily a loop---are a generalization of the
{\it centre of mass coordinates} defined by the weighted sum (or integral) over
the position vectors of matter distribution within the object. Assuming that
the object---which exists in $V_n$---consists of discrete events with
coordinates $X_i^\mu$, all having the same weight, we have
\be
     X^\mu = {1\oo N} \sum_i X_i^\mu
\lbl{20}
\ee
where $N$ is the number of events. The events are assigned the same weight
$1/N$ if they have the same masses. In general masses $m_i$ are different
and we have
\be
    X^\mu =  \sum_i X_i^\mu \rho_i
\lbl{20a}
\ee
where $\rho_i \equiv m_i/\sum_j m_j$ is the weight associated to the i-th event.
The centre of mass coordinates $X^\mu$
are collective {\it vector coordinates} associated with the object:
they determine the ``effective" position of the object.

Let us now generalize the concept of weight
by introducing the {\it two point weights} $\rho_{ij}$ which are functions
not of a single point, but of two points within an object. Then we can define
the collective {\it bivector coordinates} $X^{\mu \nu}$ associated with
an object:
\be
    X^{\mu \nu} =  \sum_{ij} X_{i}^\mu  X_{j}^\nu \rho_{ij}
\lbl{21}
\ee
They determine the effective 2-vector orientation
and area of the object. In other words, $X^{\mu \nu}$ determine how much
the object---with a given $\rho_{ij}$---deviates from a point like object:
they determine a fictitious
1-loop enclosing a 2-area defined by the bivector $X^{\mu \nu}$.

In general, the collective $r$-vector coordinates  $X^{\mu_1...\mu_r}$,
$r \le n$, associated with an object in an $n$-dimensional space
we define as
\be
     X^{\mu_1...\mu_r} =  \sum_{i_1,i_2,...,i_r}
     X_{i_1}^{\mu_1} X_{i_2}^{\mu_2}...X_{i_r}^{\mu_r} 
     \rho_{i_1 i_2 ...i_r}
\lbl{22}
\ee
They determine a fictitious $(r-1)$-loop which is a measure of the object's
effective $r$-volume. In particular, the considered objects may actually
be an $(r-1)$-loop, but in general it need not be: and yet, using the
definition (\ref{22}) we still associate with the object an $(r-1)$-loop,
which can be called a {\it centre of mass $(r-1)$-loop}. The corresponding
0-loop is the ordinary {\it centre of mass}.

In the usual dynamics (relativistic or non relativistic) we are used to
describing
the extended objects---which are not strings or the Dirac-Nambu-Goto branes---
as being point particles sitting in the object centre
of mass. Thus we take into account the information about the motion of the
object's centre of mass, but neglect the extension of the object.

In general, of course, the object's {\it extension} also should be included
into a dynamical description. The object's $r$-vector coordinates (\ref{22})
provide a natural means of how to take into account the fact that the
object is not a point particle but has a finite extension.

\subsection{Centre of mass polyvector}

The above description was just to introduce the concept of $r$-vector
coordinates of a physical object. We will now further formalize this
by introducing the {\it centre of mass polyvector} $X$ according to
    $$X = \rho {\underline 1} + \int \dd^n x \, \rho (x) x^\mu \gamma_\mu
    + {1\oo {2!}} \int \dd^n x \, \dd^N x' \, \rho (x,x') x^\mu x'^\mu \,
    \gamma_\mu \wedge \gamma_\nu$$
\be   \hs{2.5cm}   + \, {1\oo {3!}} \int \dd^n x \, \dd^N x' \, \dd^N x'' \,
    \rho(x,x',x'') x^\mu x^\nu x^\alpha \, \gamma_\mu \wedge \gamma_\nu
    \wedge \gamma_\alpha + ...
\lbl{M1}
\ee
Here $\rho (x)$ is the ordinary mass density, normalized according to
$\int \rho (x) \dd^n x = 1$, whilst $\rho,~\rho(x,x'),~\rho(x,x',x'')$,...,
are the corresponding 0-point, 2-point, 3-point, ..., generalizations.
Namely, we generalize the concept of mass density so that besides the usual
mass density $\rho(x)$ which is a 1-point function, we introduce also ``mass"
densities which are functions of any number of points.

We can distinguish two particular cases:

\ \ (i) Mass densities have support on a discrete set of spacetime points:
\bear
      \rho(x) &=& \sum_i \rho_i \, \delta (x - X_i) \nonumber \\
      \rho(x,x') &=& \sum_{ij} \rho_{ij} \, \delta (x - X_i) \delta (x - X_j)
      \nonumber \\             
              &\vdots& \lbl{M2}
\ear
Then the centre of mass polyvector (\ref{M1}) becomes
\be
      X = \rho {\underline 1} + \sum_i \rho_i \, X_i^\mu \gamma_\mu +
  {1\oo {2!}} \sum_{ij} \rho_{ij} \,  X_i^\mu X_j^\nu 
  \gamma_\mu \wedge \gamma_\nu
      + ...
\lbl{M3}
\ee
Since, on the other hand $X$ can be expanded according to eq.(\ref{23}) its
multivector components are given by (\ref{20})--(\ref{22}). The scalar component
is $\rho$.

\ (ii) Mass densities have support on a brane described by embedding functions
$X^\mu (\xi) \equiv X^\mu (\xi^a)$, $\mu = 0,1,...,n-1; a=0,1,...,p\,$:
\bear
     \rho(x) &=& \int \dd \xi \, \rho (\xi) \, \delta (x - X(\xi))
     \nonumber \\
     \rho(x,x') &=& \int \dd \xi \, \dd \xi' \, \rho(\xi,\xi') 
   \delta (x - X(\xi)) \delta (x' - X(\xi')) \nonumber \\
      &\vdots& \lbl{M4}
\ear
Then the centre of mass polyvector (\ref{M1}) is
\be
     X = \rho \, {\underline 1} + \int \dd \xi \, \rho (\xi) \, X^\mu  (\xi)
     \gamma_\mu + {1\oo {2!}} \int \dd \xi \, \dd \xi' \, \rho(\xi,\xi') \, 
   X^\mu (\xi) X^\nu (\xi')\gamma_\mu \wedge \gamma_\nu + ...
\lbl{M5}
\ee

(iii) Mass densities have support on a set of branes of various 
dimensionalities, each brane being described by the embedding functions
$X_i^\mu (\xi) \equiv X_i^\mu (\xi^{a_i}),~\mu = 0,1,2,...,n-1;~a_i =
0,1,2,...,p_i \, $:
\bear
     \rho(x) &=& \sum_i \int \dd \xi \, \rho_i (\xi) \, \delta (x - X_i(\xi))
     \nonumber \\
     \rho(x,x') &=& \sum_{ij} \int \dd \xi \, \dd \xi' \, \rho_{ij}(\xi,\xi') \,
   \delta (x - X_i (\xi)) \delta (x' - X_j(\xi')) \nonumber \\
      &\vdots& \lbl{M5a}
\ear
Then the centre of mass polyvector (\ref{M1}) assumes the form
\be
     X = \rho \, {\underline 1} + \sum_i \int \dd \xi \, \rho_i (\xi) 
     X_i^\mu  (\xi) \gamma_\mu + 
     {1\oo {2!}} \sum_{ij} \int \dd \xi \, \dd \xi' \, \rho_{ij} (\xi,\xi') 
   X_i^\mu (\xi) X_j^\nu (\xi')\gamma_\mu \wedge \gamma_\nu + ... 
\lbl{M6}
\ee 

For simplicity we have omitted the subscript $i$ on the $i$-th brane
parameters $\xi\equiv \xi^{a_i}$.

Suppose now that our system consists of branes whose sizes can be
neglected at large distances so that approximately we have
\be
     \rho_i (\xi) = \rho_i \delta (\xi - {\bar \xi}) \; , \qquad
     \rho_{ij} (\xi,\xi') = \rho_{ij} \delta (\xi-{\bar \xi}) 
     \delta (\xi' - {\bar \xi})
\lbl{M7}
\ee
Then the expression (\ref{M6}) can be approximated by the expression
(\ref{M3}). The latter expression (\ref{M3}) has non vanishing $r$-vector
part, $r = 2,3,...,$ when $\rho_{ij},~\rho_{ijk}$,..., are different
from zero and are not symmetric in $i,j...$, i.e., they contain the 
antisymmetric parts. Then our system is anisostropic, owing to the asymmetric
2-point, 3-point,..., mass densities $\rho_{ij},~\rho_{ijk},...$, and this
manifests itself in non vanishing multivector centre of mass coordinates
$X^{\mu \nu},~X^{\mu \nu \alpha},...$.
However, we may envisage the situation in which $\rho_{ij},~\rho_{ijk},...$
are symmetric in their indices. Then our system---neglecting the fact
that its constituents are not point particles, but branes---is {\it isotropic},
and consequently $X^{\mu \nu},~X^{\mu \nu \alpha},...$ are zero.

\setlength{\unitlength}{.5mm}
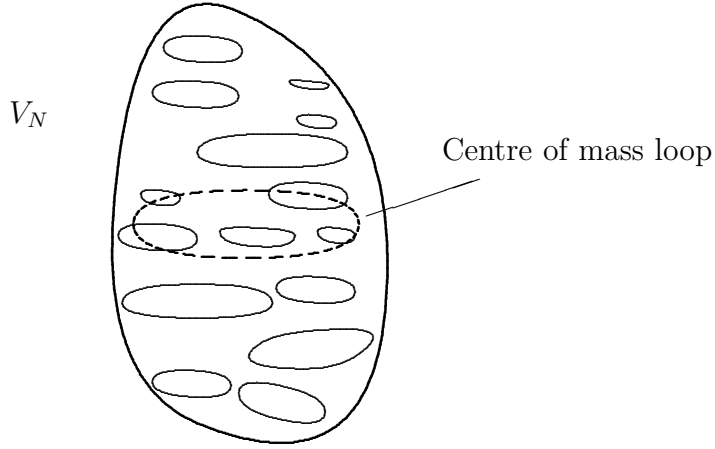
\begin{figure}[h]
\begin{picture}(150,130)(-70,-8)

\put(52,99){\closecurve(10,0, 0,4, -11,1, 0,-3)}
\put(80,20){\closecurve(15,0, 10,5, -15,0, -1,-5.5)}
\put(49,10){\closecurve(10,0, 0,4, -11,1, 0,-3)}
\put(72,5){\closecurve(0,-4, 12,-2, 0,5, -11,3)}
\put(80,60){\closecurve(10,0, 0,4, -11,1, 0,-3)}
\put(50,32){\closecurve(0,-4, 20,0, 0,5,-20,0)}
\put(70,72){\closecurve(0,-4, 20,0, 0,5,-20,0)}
\put(40,49){\closecurve(10,0, 0,4, -11,1, 0,-3)}
\put(50,87){\closecurve(11,0, 0,4, -12,1, 0,-3)}
\put(82,35){\closecurve(10,0, 0,4, -11,1, 0,-3)}
\put(66,50){\closecurve(10,0, 0,2, -10,1, 0,-3)}
\put(82,80){\closecurve(5,0, 0,2, -5.5,1, 0,-1.5)}
\put(80,90){\closecurve(5,0, 0,1, -5.5,1, 0,-1)}
\put(40,60){\closecurve(5.5,0, 0,2, -5,1, 0,-2)}
\put(87,50){\closecurve(5,0, 0,2, -5,1, 0,-2)}

\put(95,55){\line(3,1){30}}

\thicklines

\put(0,80){$V_N$}
\put(115,70){Centre of mass loop}

\closecurve(50,0, 100,30, 60,110, 28,60)

\renewcommand{\xscale}{1.}
\renewcommand{\yscale}{.3}

\curvedashes[1mm]{0,1,1}
\put(63,53){\closecurve(30,0, 0,30, -30,0, 0,-30)}

\end{picture}

\caption{If a physical object is a system of 1-loops with a prefered 
orientation, then we can associate with the system an ``effective" or 
{\it centre of mass 1-loop}.}

\end{figure} 

Now suppose that we look at our system more closely and take into account
that it consists of branes. Even if $\rho_{ij} (\xi,\xi')$ in eq.\,(\ref{M6})
is symmetric in the indices $i,j$, the bivector term in (\ref{M6}) can be
different from zero if there is antisymmetry in the parameters $\xi,\xi'$.
For instance, if we assume that
\be
    \rho_{ij} (\xi,\xi') = \delta_{ij} \rho_j (\xi,\xi')
\lbl{M8}
\ee
where $\rho_i (\xi,\xi') = - \rho_i (\xi',\xi)$, then the bivector part
of the centre of mass polyvector (\ref{M6}) then becomes
\be
    \langle X \rangle_2 = {1\oo 2} X^{\mu \nu} \gamma_\mu \wedge \gamma_\nu
    = {1\oo 2} \sum_i \int \dd \xi \, \dd \xi' \, \rho_i (\xi,\xi')
    X_i^\mu (\xi) X_i^\nu (\xi') \gamma_\mu \wedge \gamma_\nu
\lbl{M9}
\ee
from which we have
\be
     X^{\mu \nu} = {1 \oo {2}} \sum_i \int \dd \xi \, \dd \xi' \, 
     \rho_i (\xi,\xi') (X_i^\mu (\xi) X_i^\nu (\xi') -  
     X_i^\mu (\xi') X_i^\nu (\xi))
\lbl{M10}
\ee

Suppose that all the branes within the system are closed 1-branes, i.e.,
1-loops. If there are many branes and if they are randomly oriented, 
then the sum in eq.(\ref{M10})
results in $X^{\mu \nu} \approx 0$. But, if the branes have a 
prefered orientation,
then the centre of mass bivector coordinates $X^{\mu \nu}$ differ from
zero and they determine an ``effective" or {\it centre of mass 1-loop}
associated with our system of loops (branes) (see Fig.\,1).

\subsection{A special case: the extended object is a single loop}

Let us now consider a special case in which our extended object is a single
1-loop parametrized by $\xi$. Eq. (\ref{M10}) then reads
\be
    X^{\mu \nu} = {1 \oo 2} \int \dd \xi \, \dd \xi' \, \rho (\xi,\xi')
    (X^\mu (\xi) X^\nu (\xi') - X^\mu (\xi') X^\nu (\xi))
\lbl{M11}
\ee
where the integral runs over the (closed) loop so that the symbol $\int$
can be replaced by $\oint$.
We shall now demonstrate that for a suitably chosen $\rho (\xi ,\xi')$
the expression (\ref{M11}) for 
$X^{\mu \nu}$ is identical to the expression (\ref{15})
which determines the area enclosed by the loop. We shall consider two possible
choices of $\rho(\xi,\xi')$.

According to the {\it first choice} we set
\be
    \rho(\xi,\xi')  = {\p \oo {\p \xi}} \delta (\xi - \xi')
\lbl{B1}
\ee
Inserting this into eq.(\ref{M11}) we obtain after integrating out $\xi'$
the expression
\be
      X^{\mu \nu} = {1\oo 2} \oint \dd \xi \, \left ( X^\mu (\xi)
      {{\p X^\nu (\xi)}\oo {\p \xi}} - X^\nu (\xi)
      {{\p X^\mu (\xi)}\oo {\p \xi}}  \right )
\lbl{B2}
\ee
which---according to the Stokes theorem---is equivalent to the expression 
(\ref{15}) for the oriented area of enclosed by the loop.

So we have shown that a loop can be considered as an extended object,
with the two point density (\ref{B1}), whose {\it centre of mass
bivector coordinates} $X^{\mu \nu}$ ---calculated according to eq.(\ref{M11})
(which in turn is a part of the general formula (\ref{M1}))--- are just the 
projections of the oriented area enclosed by the loop. By this we have
checked the consistency of the definition (\ref{M1}) for the
centre of mass bivector coordinates $X^{\mu \nu}$. The analog is expected
to be true for higher multivector coordinates $X^{\mu \nu \alpha ...}$.

According to {\it the second choice} we can set
\be
     \rho (\xi, \xi') = {1\oo 2} \, \epsilon (\xi,\xi')
\lbl{B3}
\ee
where $\epsilon (\xi,\xi') = - \epsilon (\xi', \xi) = 1$. Then in the
case in which the loop is a circle we have
\bear
    X^{\mu \nu} &=& {1\oo 4} \int \dd \varphi \dd \varphi' \, 
    \epsilon (\varphi,\varphi')
    (X^\mu (\varphi) X^\nu (\varphi') - X^\nu (\varphi) X^\mu (\varphi') )
    \nonumber \\
     &=& {1\oo 2} \int_0^{2 \pi} \dd \varphi \int_\varphi^{2 \pi} \dd \varphi'
    (X^\mu (\varphi) X^\nu (\varphi') - X^\nu (\varphi) X^\mu (\varphi') )
\lbl{B4}
\ear
Taking the parametric equation of the circle, namely $X^1 = r \, {\rm cos} \, 
\varphi$, $X^2 = r \, {\rm sin} \, \varphi$, we find that

\be
   X^{12} = {r^2\oo 2} \int_0^{2 \pi} \dd \varphi \int_\varphi^{2 \pi}   
      \dd \varphi' \, ({\rm cos}\, \varphi \, {\rm sin} \, \varphi' -
      {\rm sin} \, \varphi \, {\rm cos} \, \varphi' )
      = {r^2\oo 2} \int_0^{2 \pi} \dd \varphi (- {\rm cos} \, \varphi + 1)
      = \pi r^2
\lbl{B5}
\ee
which is just the area of the circle. It would be interesting to investigate
what eq.(\ref{B4}) would give for an arbitrary closed curve $X^\mu (\varphi)$.
 
\section{Relativity in $C$-space}

\subsection{Kinematics}

All $r$-vector coordinates of a physical object are in principle different
from zero, and they all have to be taken into account. Vectors $X^\mu 
\gamma_\mu$, bivectors ${1\oo {2!}} X^{\mu \nu} \gamma_\mu \wedge 
\gamma_\nu$, and
in general $r$-vectors (or multivectors) ${1\oo {r!}} X^{\mu_1...\mu_r}
\gamma_{\mu_1} \wedge ... \wedge \gamma_{\mu_r}$ they all enter the
description of an object. 
\paragraph{Compact notation} A coordinate polyvector can be written as
\be
    X = {1\oo {r!}} \sum_{r=0}^N X^{\mu_1...\mu_r} \gamma_{\mu_1} \wedge
    ...\wedge \gamma_{\mu_r} \equiv X^A E_A
\lbl{24a}
\ee
Here we use a compact notation in which $X^A \equiv X^{\mu_1...\mu_r} $ are
coordinates, and $E_A \equiv  \gamma_{\mu_1} \wedge...\wedge \gamma_{\mu_r} $,
$\mu_1 < \mu_2 < ... <\mu_r$,
basis vectors of the $2^n$-dimensional {\it Clifford algebra} of spacetime.
The latter algebra of spacetime positions and corresponding higher grade
objects, namely oriented $r$-areas, is a manifold which is more general than
spacetime. In the literature such a manifold has been named {\it pandimensional
continuum} \ci{Pezzaglia} or {\it Clifford space} or $C$-{\it space} 
\ci{Castro}.

The infinitesimal element of position polyvector (\ref{11}) is
\be
      \dd X = {1\oo {r!}} \sum_{r=0}^N \dd X^{\mu_1...\mu_r} 
      \gamma_{\mu_1} \wedge
    ...\wedge \gamma_{\mu_r} \equiv \dd X^A E_A
\lbl{24b}
\ee
We will now calculate its square. Using the definition for the {\it scalar
product} of two polyvectors $A$ and $B$
\be
      A * B = \langle A B \rangle_0
\lbl{13a}
\ee
where $\langle \; \rangle_0$ means the scalar part of the geometric product
$A B$, we obtain
\be
     |\dd X |^2 \equiv \dd X^{\dagger} * \dd X = \dd X^A \, \dd X^B
     G_{AB} = \dd X^A \, \dd X_A
\lbl{24c}
\ee
Here
\be
       G_{AB} = E_A^{\dagger} * E_B
\lbl{24d}
\ee
is the $C$-space metric and $A^{\dagger}$ the reverse\footnote{Reversion or
{\it hermitian conjugation} is the operation \ci{Hestenes} which reverses 
the order of all products of vectors in a decomposition of a polyvector $A$.
For instance, $(\gamma_1 \gamma_2 \gamma_3)^\dagger = \gamma_3 \gamma_2 
\gamma_1$. }
of a polyvector $A$.

For example, if the indices assume the values $A=\mu, \; B=\nu$, we have
\be
     G_{\mu \nu} = \langle e_\mu e_\mu \rangle_0 = e_\mu \cdot e_\nu =
     g_{\mu \nu}
\lbl{24e}
\ee
If $A = [\mu \nu]$, $B=[\alpha \beta]$
\bear
      G_{[\mu \nu][\alpha \beta]} &=& \langle (e_\mu \wedge e_\nu)^\dagger 
      (e_\alpha \wedge e_\beta \rangle_0 = 
      \langle (e_\mu \wedge e_\nu)^\dagger \cdot (e_\alpha
      \wedge e_\beta \rangle_0 \nonumber \\
       &=& (e_\mu \cdot e_\alpha)(e_\nu \cdot e_\beta)
      - (e_\nu \cdot e_\alpha)(e_\mu \cdot e_\beta) = g_{\mu \alpha}
      g_{\nu \beta} - g_{\nu \alpha} g_{\mu \beta}
\lbl{24f}
\ear
If $A=\mu$, $B= [\alpha \beta]$
\be
      G_{\mu [\alpha \beta]} = \langle e_\mu (e_\alpha \wedge e_\beta)
      \rangle_0 = 0
\lbl{24g}
\ee

Explicitly we have
\bear
     |\dd X |^2 &=& {1\oo {r!}} \sum_{r=0}^N \dd  X^{\mu_1...\mu_r}   
     \dd  X_{\mu_1...\mu_r} \nonumber \\
      &=& \dd \sigma^2 + \dd X^\mu \dd X_\mu + {1\oo {2!}}
     \dd X^{\mu_1 \mu_2} \dd X_{\mu_1 \mu_2} + ... + {1\oo {n!}} 
     \dd X^{\mu_1 ... \mu_n} \dd X_{\mu_1 ... \mu_n}
\lbl{24h}
\ear

In $C$-space the usual points, lines, surfaces, etc., are all described
on the same footing and can be transformed into each other by rotations
in $C$-space (polydimensional rotations):
\be
     X'^A = {L^A}_B X^B
\lbl{24A} 
\ee
These are generalizations of the usual Lorentz transformations.
They mix multivectors of different grades. For instance, a pure vector 
can acquire a bivector component, so that after the transformation
it is a superposition of a vector and a bivector. This means that 
a point with coordinates $X^A = (0, x^\mu, 0, 0, 0)$ can become a loop with 
coordinates $X'^A = (0, x'^\mu, x'^{\mu \nu}, 0, 0)$.

The generalized Lorentz transformations, i.e., the Lorentz transformations
in $C$-space, include the ordinary Lorentz boosts which connect the
reference systems in relative translational motion. In addition they also
include the boost which connect the reference frames in relative
dilatational motion. All experiences that we have from the relativity in
Minkowski space are directly applicable to $C$-space. The line element,
the action, the transformations between reference frames have the same
form, only the set of variables (and its interpretation) is different.
Instead of $x^\mu$ we have $X^A$, which are the holographic projections 
of the centre of mass $(r-1)$-loops.

\subsection{Dynamics} 

The dynamical variables of our physical object are given by a polyvector
$X$. The {\it action} is a generalization of the point particle action:
\be
      I = \kappa \int \dd \tau \, ({\dot X}^{\dagger} * {\dot X})^{1/2} 
        = \kappa \int \dd \tau \, ({\dot X}^A {\dot X}_A )^{1/2}
\lbl{25}
\ee
where $\kappa$ is a constant, playing the role of ``mass" in $C$-space,
and $\tau$ is an arbitrary parameter.

The theory so obtained is an extension of the ordinary special relativity.
We shall call it {\it extended relativity} or {\it relativity in $C$-space}.

The equation of motion resulting from (\ref{25}) is

\be
    {\dd \oo {\dd \tau}} \left ( {{{\dot X}^A}\oo 
    {\sqrt{{\dot X}^B {\dot X}_B}}} \right ) = 0
\lbl{26}
\ee
Taking ${\dot X}^B {\dot X}_B = {\rm constant} \neq 0$ we have that 
${\ddot X}^A= 0$, so that $X^A (\tau)$ is a straight worldline in $C$-space.
The components $X^A$ then change linearly with the parameter $\tau$.

Let us consider more
closely what does it mean physically that the multivector components
are linear functions of $\tau$. This implies that the holographic
projections $X^A \equiv X^{\mu_1 ...\mu_r}$ of the centre of mass $(r-1)$-loops are
linear functions of $\tau$. This means that  the orientation and the area
of the $r$-surface enclosed by the $(r-1)$-loop can increase or decrease.
In particular, the initial conditions can be such that the orientation
remains the same and only the area changes. In such a case we have a
pure {\it dilatational motion} (without rotational motion) of the
$(r-1)$-loop. The loop changes its {\it scale} (i.e., its size, extension)
during its motion (Fig.\,2). The dilatational degree of freedom 
(briefly, {\it scale}) is
encoded in the holographic coordinates $X^{\mu_1 ...\mu_r}$ for
$r\ge 2$. The theory here suggests that {\it scale} is a degree of freedom
in the analogous way as the centre of mass position is a degree of freedom.
Namely, by definition $X^{\mu \nu ...}$ are degrees of freedom, and we have
seen that they determine the effective extension (size or scale) of the
object.

\setlength{\unitlength}{.6mm}
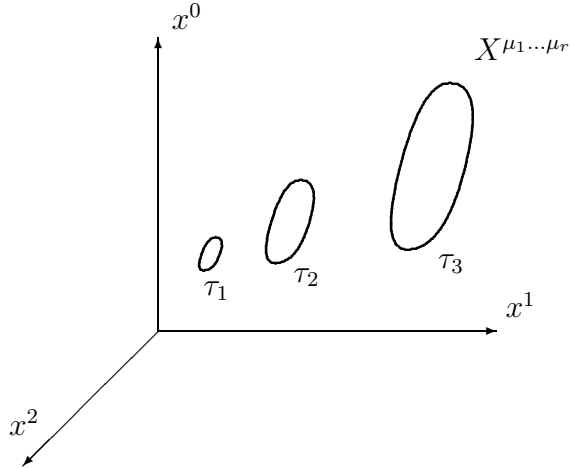
\begin{figure}[h]
\begin{picture}(150,100)(-90,-30)

\put(0,0){\vector(1,0){75}}
\put(0,0){\vector(0,1){65}}
\put(0,0){\vector(-1,-1){30}}

\put(77,3){$x^1$}
\put(-33,-23){$x^2$}
\put(3,67){$x^0$}

\put(10,8){$\tau_1$}
\put(30,11){$\tau_2$}
\put(62,14){$\tau_3$}
\put(70,60){$X^{\mu_1...\mu_r}$}

\thicklines

\renewcommand{\xscale}{.24}
\renewcommand{\yscale}{.2}
\put(12,17){\closecurve(-8,-18, 7,0, 3,19, -10,0)}

\renewcommand{\xscale}{.5}
\renewcommand{\yscale}{.5}
\put(30,24){\closecurve(-8,-18, 7,0, 3,19, -10,0)}

\renewcommand{\xscale}{.85}
\renewcommand{\yscale}{1}
\put(62,36){\closecurve(-8,-18, 7,0, 3,19, -10,0)}

\end{picture}

\caption{In general an $(r-1)$-loop changes with $\tau$. Its 
{\it orientation} and {\it size} increases or decreases. The picture
illustrates the {\it dilatational motion} of a loop.}

\end{figure} 

So we have discovered that {\it scale} as a degree of freedom of an object is 
inherent in the Clifford algebra extension of the usual point particle
kinematics and dynamics. The dynamics in $C$-space is more general than
it is usually assumed. In the usual dynamics the scale of a free rigid object is
fixed; it cannot change. Only the object's centre of mass position can change.
In the $C$-space dynamics the scale of a free object can change. The scaled
object remains similar to the original object, only all the distances
and other scale dependent quantities within or related to the object are
changed by a scale factor. Such a scale change should be understood as
to be applicable to all sorts of objects and its constituents,  
the atoms, nuclei, etc. 
The theory predicts {\it scaled atoms} with scaled spectral lines.
A possible objection, namely that quantum mechanics does not allow this,
is not valid here. The ordinary quantum mechanics takes into account
the position $X^\mu$ and position dependent forces between the objects
(i.e., nuclei and electrons) constituting the atoms and so it comes out
that atoms have fixed sizes. If we generalize quantum mechanics so to
include the holographic coordinates $X^{\mu \nu ...}$ into the description,
the situation is different. Scale then becomes a degree of freedom and has
to be quantized as well. Like position, the scale of a {\it free} object
can be arbitrary, whilst the scale of a bound object has discrete quantum
values. If the atom as a whole is a free object, its centre of mass
position can be arbitrary. Similarly, its {\it scale} can be arbitrary.
What is not arbitrary is relative positions of electrons within the atom:
they are determined by the solutions of the Schr\" odinger equation
(generalized to $C$-space). The translational and dilatational motion of the
atom as a whole, i.e., how the centre of mass coordinate $X^\mu$ and the
generalized centre of mass coordinates $X^{\mu \nu ...}$ move, also is
determined by the Schr\" odinger equation in $C$-space. We shall not go
into a more detailed description here (see refs. \ci{Pavsic1,Pavsic:1976ta}).

The objects's 4-volume $X^{0123}$ also changes with $\tau$. With increasing 
$\tau$ an arbitrary object in $V_4$ can become more and more similar to 
{\it world line} (Fig.\,3).

\setlength{\unitlength}{.7mm}
\begin{figure}[h]
\begin{picture}(150,100)(-90,-30)

\put(0,0){\vector(1,0){75}}
\put(0,0){\vector(0,1){65}}
\put(0,0){\vector(-1,-1){30}}

\put(77,3){$x^1$}
\put(-33,-23){$x^2$}
\put(3,67){$x^0$}

\put(-22,40){$V_4$}

\put(11,23){$\tau_1$}
\put(30,25){$\tau_2$}
\put(46,36){$\tau_2$}
\put(63,60){$\tau_4$}

\thicklines

\renewcommand{\xscale}{.18}
\renewcommand{\yscale}{.18}
\put(14,17){\closecurve(0,-24, 20,0, 0,19, -22,0)}

\renewcommand{\xscale}{.30}
\renewcommand{\yscale}{.40}
\put(30,15){\closecurve(-8,-18, 7,0, 3,19, -10,0)}

\put(45,16){
\begin{rotate}{-10}
\renewcommand{\xscale}{.17}
\renewcommand{\yscale}{.36}
\closecurve(0,-36, 9,-34, 11,0, 10,35, 8,43, 0,45, -8,43, -10,35, -11,0, -9,-34)
 \end{rotate} }

\put(60,16){
\begin{rotate}{-10}
\renewcommand{\xscale}{.08}
\renewcommand{\yscale}{1.2}
\closecurve(0,-36, 9,-34, 11,0, 10,35, 8,43, 0,45, -8,43, -10,35, -11,0, -9,-34)
 \end{rotate} }

\end{picture}

\caption{Illustration of 4-volume changes as a result of the dynamics in $C$-space.
An object's 4-volume increases when its 4-vector speed is different from
zero and positive. If the space-like components of its 3-vector and
2-vector velocity are zero or negative, the 4-volume can increase only
if the object become more and more elongated in the time-like direction}

\end{figure}
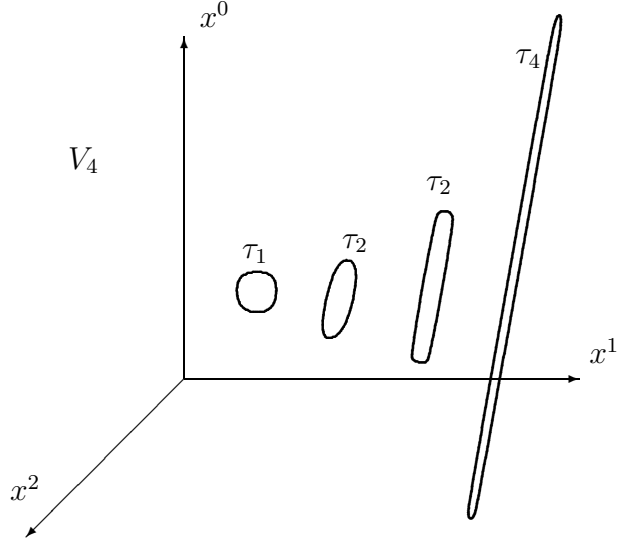 

So the physics in $C$-space predicts long, but not infinite, worldlines.
This has consequences for the electromagnetic interaction.

\subsection{The electromagnetic field}

Although a complete treatment of the minimal coupling and gauge fields in
$C$-space remains to be elaborated\footnote{This is one of our next
projects \ci{Aurilia}.}
we anticipate here that the electromagnetic potential $A^{\mu}$ around
a finite worldline is
\be
     A^\mu (x) = \int_{-\ell/2}^{\ell/2} e \, \delta [(x - X(\lambda))^2]
     {\dot X}^\mu (\lambda) \, \dd \lambda
\lbl{3.1}
\ee
Here $X^\mu (\lambda)$ are embedding functions of the worldline, paramterized
by $\lambda$, $e$ the electric charge and $x^\mu$ the coordinates of any point
in the embedding spacetime (Fig.\,4).

\setlength{\unitlength}{.6mm}
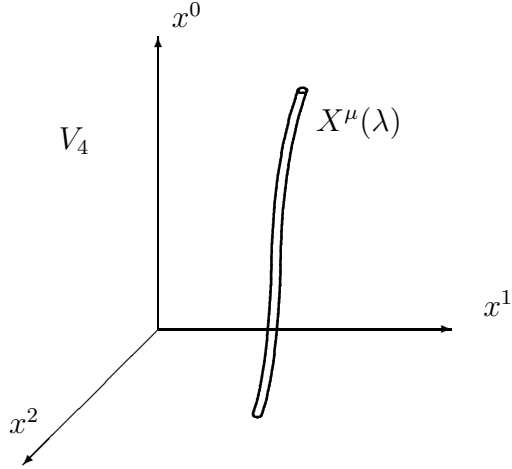
\begin{figure}[h]
\begin{picture}(150,100)(-90,-30)

\put(0,0){\vector(1,0){65}}
\put(0,0){\vector(0,1){65}}
\put(0,0){\vector(-1,-1){30}}

\put(72,3){$x^1$}
\put(-33,-23){$x^2$}
\put(3,67){$x^0$}

\put(-22,40){$V_4$}
\put(35,44){$X^\mu (\lambda)$}

\thicklines

\renewcommand{\xscale}{1.}
\renewcommand{\yscale}{1.}
\put(26,11){
\renewcommand{\xscale}{1.}
\renewcommand{\yscale}{.6}
\curve(-5,-50, -4,-50.8, -3,-50)}

\renewcommand{\xscale}{1.}
\renewcommand{\yscale}{1.2}

\put(-60,-50){
\qbezier(81,26)(84,33)(85,50)
\qbezier(85,50)(85,72)(91,86) }

\put(-58,-50){
\qbezier(81,26)(84,33)(85,50)
\qbezier(85,50)(85,72)(91,86) }
\put(32,53){
  \renewcommand{\xscale}{.1}
  \renewcommand{\yscale}{.06}
  \closecurve(10,0, 0,10, -10,0, 0,-10) }

\end{picture}

\caption{A finite, electrically charged, worldline-like object ${\cal O}$
of a finite length $\ell$, produces at an arbitrary point $x^\mu$ the 
electromagnetic potential $A^\mu (x)$ which deviates from the ordinary
electromagnetic potential. The deviation becomes smaller and smaller
with increasing $\ell$.}

\end{figure}

The ordinary electromagnetic interaction is obtained in the limit of
infinitly long worldlines ($\ell \rightarrow \infty$). In this theory
electromagnetic interaction depends on $\ell$. Since $\ell$
changes with
$\tau$, electromagnetic interaction also changes. It remains as a
future project to elaborate this and calculate numerically how
electromagnetic interaction changes, and compare the result with the
recent claims based, on astrophysical data, that the fine structure
constant $\alpha$ has changed during the evolution of the Universe \ci{alpha}.

\subsection{Estimation of holographic speeds}

From the action (\ref{25}) we find the following expression for
the momentum in $C$-space:
\be
     P_A = {{\kappa {\dot X}_A}\oo {({\dot X}^B {\dot X}_B)^{1/2}}}
\lbl{3.2}
\ee
When the denominator in eq.(3.2) is zero the momentum becomes infinite.
We shall now calculate the speed at which this happens. This will be the {\it
maximum speed} that an object accelerating in $C$-space can reach\footnote{Although
an initially slow object cannot accelerate beyond that speed limit, this does
not automatically exclude the possibility that fast objects traveling at
a speed above that limit may exist. Such objects are $C$-space analog
of tachyons. All the well known objections against tachyons should be
reconsidered for the case of $C$-space before we could say for sure that
$C$-space tachyons do not exist as freely propagating objects.}.

Vanishing of ${\dot X}^B {\dot X}_B$ is equivalent to vanishing of the
$C$-space line element
\be
    \dd X^A \dd X_A = \dd \sigma^2 + \left ( {{\dd X^0}\oo L} \right )^2
    - \left ( {{\dd X^1}\oo L} \right )^2 + \left ( {{ \dd X^{12}}\oo
    L^2} \right )^2 - \left ( {{\dd X^{123}}\oo L^3 } \right )^2
    - \left ( {{\dd X^{0123}}\oo L^4} \right )^2 + ... = 0
\lbl{3.3}
\ee
where by ``..." we mean the terms with the remaining components such as
$X^2$, $X^{01}$, $X^{23}$,..., $X^{012}$, etc.. In eq. (\ref{3.3}) we introduce
a length parameter $ L$. This is necessary, since $X^0 = c t$ has
dimension of length, $X^{12}$ of length square, $X^{123}$ of length
to the third power, and $X^{0123}$ of length to the forth power.
It is natural to assume that $L$ is the {\it Planck length},
that is $L = 1.6 \times 10^{-35} m$.

Let us assume that the coordinate time $t = X^0/c$ is the parameter with 
respect to which we  define the speed $V$ in $C$-space. So we have
\be
    V^2 = - \left (L {{\dd \sigma}\oo {\dd t}} \right )^2 + \left ( {{\dd X^1}
    \oo {\dd t}} \right )^2 - \left ( {1\oo L} {{\dd X^{12}}\oo {\dd t}}
    \right )^2 + \left ({1\oo L^2} {{\dd X^{123}}\oo {\dd t}} \right )^2
    + \left ( {1\oo L^3} {{\dd X^{0123}}\oo {\dd t}} \right )^2 - ...
\lbl{3.4}
\ee
The maximum speed is given by
\be
       V^2 = c^2
\lbl{3.5}
\ee

First, we see that the maximum speed squared $V^2$ contains not only the
components of the 1-vector velocity $\dd X^1/\dd t$, as it is the case
in the ordinary relativity, but also the multivector components such as
$\dd X^{12}/\dd t$, $\dd X^{123}/\dd t$, etc..

The following special cases when only certain components of the velocity in
$C$-space are different from zero, are of particular interest:
\begin{itemize}
  
\item[(i)] Maximum 1-vector speed
   $${{\dd X^1}\oo {\dd t}} = c = 3.0 \times 10^8 m/s $$
   
\item[(ii)] Maximum 3-vector speed
   \bear
   &&{{\dd X^{123}}\oo {\dd t}} = L^2 c = 7.7 \times 10^{-62} m^3/s
   \nonumber \\
   && {{\dd \root 3 \of {X^{123}} }\oo {\dd t}} = 4.3 \times 10^{-21} m/s
   \quad \mbox{(diameter speed)} \nonumber
   \ear

\item[(iii)] Maximum 4-vector speed
   \bear
   &&{{\dd X^{0123}}\oo {\dd t}} = L^3 c = 1.2 \times 10^{-96} 
   {\rm m}^4/{\rm s} \nonumber \\
   && {{\dd \root 4 \of {X^{0123}} }\oo {\dd t}} = 1.05 \times 10^{-24} m/s
   \quad \mbox{(diameter speed)} \nonumber
   \ear
   
\end{itemize}   
Above we have also calculated the corresponding diameter speeds for the
illustration of how fast the obejct expands or contracts.

We see that the maximum multivector speeds are very small. The diameters
of objects change very slowly. Therefore we normally do not
observe the dilatational motion.

Because of the positive sign in front of the $\sigma$ and $X^{12},~X^{012}$,
etc., terms in the quadratic form (\ref{3.3})
there are no limits to correspondintg 0-vector, 2-vector and 3-vector
speeds. But if we calculate, for instance,
the energy necessary to excite 2-vector motion we find that it is very
high. Or equivalently, to the relatively modest energies (available at 
the surface of the Earth), the 
corresponding 2-vector speed is very small. This can be seen by calculating
the energy
\be
     p^0 = {{\kappa c^2}\oo {\sqrt{1 - {{V^2}\oo c^2}}}}
\lbl{3.6}
\ee

   \ \ (a) for the case of pure 1-vector motion by taking
$V= \dd X^1/\dd t$, and
    
    \ \ (b) for the case of pure 2-vector motion by taking
$V = \dd X^{12}/(L \dd t)$.

By equating the energies belonging to the cases (a) and (b we have
\be p^0 =
{{\kappa c^2}\oo {\sqrt{1 - \left ( {1\oo c} {{\dd X^1}\oo {\dd t}} \right )^2}}}
= {{\kappa c^2}\oo {\sqrt{1 - \left ( {1\oo {Lc}} {{\dd X^{12}}\oo 
{\dd t}} \right )^2}}}
\lbl{3.7}
\ee
which gives
\be
   {1\oo c} {{\dd X^1}\oo {\dd t}} = {1\oo {Lc}} {{\dd X^{12}}\oo {\dd t}}
   = \sqrt{1 - \left ( {{\kappa c^2}\oo p_0} \right )^2 }
\lbl{3.8}
\ee
Thus to the energy of an 
object moving  translationally at $\dd X^1 /\dd t =$1 m/s, there corresponds 
the 2-vector speed $\dd X^{12}/\dd t = L \, \dd X^1 /\dd t =
1.6 \times 10^{-35}$ m$^2$/s (diameter speed 4 $\times 10^{-18}$m/s).
This would be a typical 2-vector speed of a macroscopic object. For a
microscopic object, such as the electron, which can be accelerated close
to the speed of light, the corresponding 2-vector speed could be of the
order of $10^{-26}$ m$^2$/s (diameter speed $10^{-13}$m/s).
In the examples above we have provided rough estimations of possible
2-vector speeds. Exact calculations should treat concrete situations
of collisions of two or more objects, assume that not only 1-vector, but
also 2-vector, 3-vector and 4-vector motions are possible, and take into
account the conservation of the polyvector momentum $P_A$.

{\it Maximum 1-vector speed, i.e., the usual speed, can exceed the speed
of light when the holographic components such as $\dd \sigma/\dd t$,
$\dd X^{12}/\dd t$, $\dd X^{012}/\dd t$, etc.,  are different from zero.}
This can be immediately 
verified from eqs. (\ref{3.4}),(\ref{3.5}). The speed of light is no longer 
such a strict barrier as it appears in the ordinary theory of relativity 
in $M_4$. In $C$-space a particle has extra degrees of freedom, besides the 
translational degrees of freedom. The scalar, $\sigma$, the bivector,
$X^{12}$ (in general, $X^{rs}, ~r,s = 1,2,3$) and the threevector, $X^{012}$
(in general, $X^{0rs}, ~~r,s = 1,2,3$), contributions to the $C$-space
quadratic form (\ref{3.3}) have positive sign, which is just opposite to the
contributions of other components, such as $X^r,~X^{0r},~X^{rst},
X^{\mu \nu \rho \sigma}$. Because the quadratic form has both $+$ and 
$-$ signs, the absolute value of the 3-velocity $\dd X^r/\dd X^0$ can
be greater than $c$.

\section{From the ``block universe" in $C$-space to the relativistic
dynamics and evolution in Minkowski spacetime}

Extended relativity, i.e., the relativity in $C$-space, is based
on the action (\ref{25}). An equivalent action is
\be
    I[X^A,\lambda] = {1 \oo 2} \int \dd \tau \left (
    {{{\dot X}^A {\dot X}_A}\oo \lambda} +\lambda \kappa^2\right )
\lbl{4.1}
\ee
which is a functional of the variables $X^A \equiv X^{\mu_1 ... \mu_r}$
and a scalar Lagrange multiplier $\lambda$. Variation of (\ref{4.1}) with
respect to $\lambda$ gives
\be
       {\dot X}^A {\dot X}_A = \lambda^2 \kappa^2
\lbl{4.2}
\ee
After inserting the latter relation into the action (\ref{4.1}) we obtain
the action (\ref{25}), by which we have a confirmation that both actions are
classically equivalent.
    
In the usual theory of relativity we often choose the parameter $\tau$ such that
it is equal to the coordinate $X^0 \equiv t$. In the extended relativity
we may choose parameter $\tau$ equal to the variable $s$ of the pseudoscalar
part of $X$ (see eq.(\ref{24})),
that is, we may put $\tau = s$. By doing, so we obtain a reduced theory in
which all
$X^A$, except $s$, are independent. So it was found \ci{Pavsic1} 
that the elegant 
Stueckelberg theory is naturally embedded in the $C$-space action 
(\ref{25}) or (\ref{4.1}). Here also lies a clue to a natural resolution
of the ``problem of time" in quantum gravity (see e.g. \cite{Pavsic2, Carlini}).

The discovery that we obtain the unconstrained Stueckelberg action in $M_4$
from the constrained action in $C$-space is so important that it is worth
to discuss it here again. In refs. \ci{Pavsic1} I started from a yet another
equivalent action, namely the phase space or first order action in $C$-space,
\be
     I[X^A,P_A,\lambda] = \int \dd \tau \left ( P_A {\dot X}^A -
     {\lambda \oo 2} (P^A P_A - \kappa^2) \right )
\lbl{4.2a}
\ee
which is a functional not only of the coordinate variables $X^A$, but
also of the canonically conjugate momenta $P_A$. In this paper, instead,
I will start first from the action (\ref{4.1}) and then also directly
from the action (\ref{25}).

Let us now take the dimension of spacetime $n=4$. Then the velocity
polyvector reads
\be
    {\dot X} = {\dot X}^A E_A = {\dot \sigma } {\underline 1}
    + {\dot x}^\mu \gamma_\mu + \mbox{${1\oo 2}$} {\dot x}^{\mu \nu}
    \gamma_\mu \wedge \gamma_\nu + {\dot \xi}^\mu I \gamma_\mu
    +{\dot s} I
\lbl{4.3}
\ee
Further, let us consider a special case in which
\be
    {\dot \sigma} = 0 \quad , \qquad {\dot x}^{\mu \nu} = 0 \quad ,
     \qquad {\dot \xi}^\mu = 0
\lbl{4.4}
\ee
so that the velocity polyvector and its square are simply
\be
    {\dot X} = {\dot x}^\mu \gamma_\mu + {\dot s} I
\lbl{4.5}
\ee
\be
    |{\dot X}|^2 = {\dot X}^A {\dot X}_A = {\dot x}^{\mu}{\dot x}_\mu  - 
    {\dot s}^2 
\lbl{4.6}
\ee
The actions (\ref{25}) and (\ref{4.1}) then assume the simplified forms
\be
    I[x^\mu,s] = \kappa \int \dd \tau \, \sqrt{{\dot x}^{\mu} {\dot x}_\mu- 
    {\dot s}^2}
\lbl{4.7}
\ee
and
\be
     I[x^\mu,s,\lambda] = {1 \oo 2} \int \dd \tau \, \left (
     {{{\dot x}^{\mu} {\dot x}_\mu - {\dot s}^2}\oo \lambda} +\lambda \kappa^2
     \right )
\lbl{4.8}
\ee

Let us first consider the Howe-Tucker-like \ci{Howe-Tucker} action (\ref{4.8}). 
The corresponding equations of motion are
\bear
      && \delta x^\mu \; : \qquad {\dd \oo {\dd \tau}} \left ( {{
      {\dot x}^\mu}\oo \lambda} \right ) = 0 \lbl{4.9} \\
      && \delta s \; : \; \qquad {\dd \oo {\dd \tau}} \left ( {{
      {\dot s}}\oo \lambda} \right ) = 0 \lbl{4.10} \\
      && \delta \lambda \; : \; \qquad \sqrt{{\dot x}^{\mu} - {\dot s}^2}
      = \lambda \kappa \lbl{4.11}
\ear
      
 Using eq.(\ref{4.10}) we have
\be      
   {\dd \oo {\dd \tau}} \left ({{
      {\dot s}s}\oo \lambda} \right ) = {{ {\dot s}^2}\oo \lambda}
\lbl{4.12}
\ee
Inserting the latter relation into the action (\ref{4.8}) we find
\be
    I[x^\mu,s,\lambda] = {1 \oo 2} \int \dd \tau \, \left [
    {{{\dot x}^\mu {\dot x}_\mu }\oo \lambda} + \lambda \kappa^2 - 
    {\dd \oo {\dd \tau}} \left ( {{
      {\dot s}s}\oo \lambda} \right ) \right ]
\lbl{4.13}
\ee
We will now use the fact that the Lagrange multiplier $\lambda$ can be an
arbitrary function of $\tau$: choice of $\lambda$ is related to choice of
parametrization, i.e., choice of the parameter $\tau$. Let us
choose
\be
   \lambda = \Lambda  \; , ~ \mbox{\rm i.e. ,} \qquad 
   \sqrt{{\dot x}^\mu {\dot x}_\mu- {\dot s}^2} = \Lambda \kappa
\lbl{4.14}
\ee
where $\Lambda$ is a fixed constant, and insert (\ref{4.14}) into (\ref{4.13}).
Omitting the total derivative term (which does not influence the equations of 
motion) we obtain
\be
     	I[x^\mu] = {1\oo 2} \int \dd \tau \, \left ( 
     	{{{\dot x}^\mu {\dot x}_\mu }\oo \Lambda} + \Lambda \kappa^2 \right )
\lbl{4.15}
\ee
which is just the well known {\it Stueckelberg action} in which all
$x^\mu$ are independent.

The equation of motion derived from the Stueckelberg action (\ref{4.15})
is
\be
    {\dd \oo {\dd \tau}} \left ( {{{\dot x}^\mu}\oo \Lambda} \right ) = 0
\lbl{4.16}
\ee
and they are identical to the $x^\mu$ equations of motion (\ref{4.9})
derived from the action (\ref{4.8}) in the presence of the ``gauge" condition
(\ref{4.14}). Eq. (\ref{4.16}) implies that the momentum $p_\mu =
{\dot x}_\mu/\Lambda$ and its square $p^\mu p_\mu$ are constant of motion.

The relation (\ref{4.14}) can be written in the form
\be
      \sqrt{\dd x^\mu \dd x_\mu - \dd s} = \Lambda \kappa \, \dd \tau
\lbl{4.14a}
\ee
which says that the parameter $\tau$ is equal to the length of the world line
in $C$-space.

In order to further investigate the relation (\ref{4.14})
let us consider the $s$ equation of motion (\ref{4.10}) from which we find
\be
    {{\kappa {\dot s}}\oo {\sqrt{{\dot x}^\mu {\dot x}_\mu- {\dot s}^2}}} =
    {1\oo C}
\lbl{4.17}
\ee
where $C$ is a constant of integration. From (\ref{4.17}) and (\ref{4.14})
we have the relation
\be
    {\Lambda \oo C} = {{\dd s}\oo {\dd \tau}}
\lbl{4.18}
\ee
which says that the parameter $\tau$ is proportional to the variable $s$.
By inserting (\ref{4.18}) into the action (\ref{4.15}) we find
\be
    I[x^\mu] = {1\oo 2} \int \dd s \, \left ( {1\oo C} 
    {{\dd x^\mu}\oo {\dd s}}
    {{\dd x_\mu}\oo {\dd s}} + C \kappa^2 \right )
\lbl{4.19}
\ee
The evolution parameter in (\ref{4.19}) is thus just the extra variable
$s$ entering the polyvector (\ref{4.3}).

Instead of using
the Howe-Tucker action (\ref{4.8}) we can start directly
from the action (\ref{4.7}). The equations of motion are
\be
    {\dd \oo {\dd \tau}} \left ( {{\kappa {\dot x}^\mu}\oo
    {\sqrt{{\dot x}^\mu {\dot x}_\mu- {\dot s}^2}}} \right ) = 0
\lbl{4.19a}
\ee
\be      
    {\dd \oo {\dd \tau}} \left ( {{\kappa {\dot s}}\oo
    \sqrt{{\dot x}^\mu {\dot x}_\mu- {\dot s}^2}} \right ) = 0  
\lbl{4.19b}
\ee
which is equivalent to (\ref{4.9})--(\ref{4.11}). The canonical momenta are
\be
    p_\mu = {{\kappa {\dot x}^\mu}\oo 
    \sqrt{{\dot x}^\nu {\dot x}_\nu- {\dot s}^2}} \; , \qquad
    p_{(s)} = {{\kappa {\dot s}}\oo
    \sqrt{{\dot x}^\nu {\dot x}_\nu- {\dot s}^2}}
\lbl{4.19c}
\ee
and they satisfy the constraint
\be
     p_\mu p^\mu - p_{(s)}^2 = \kappa^2
\lbl{4.19d}
\ee
Since ${\dot s} = \dd s/\dd \tau$, the action (\ref{4.7}) can be written
in the form
\be
      I[x^\mu] = \kappa \int \dd s \, \sqrt{{{\dd x^\mu}\oo {\dd s}}
      {{\dd x_\mu}\oo {\dd s}} - 1}
\lbl{4.20a}
\ee
in which the parameter $\tau$ has disappeared and the variable $s$ takes the
role of the evolution parameter.

Alternatively, one can choose parameter $\tau$ in (\ref{4.7}) such that
${\dot s} = 1$ (i.e., $\dd s = \dd \tau$) and we obtain the reduced action
\be
    I[x^\mu] = \kappa \int \dd \tau \, \sqrt{{\dot x}^\mu {\dot x}_\mu - 1}
\lbl{4.20}
\ee
which is the same as (\ref{4.20a}).

Since the action (\ref{4.20}) is {\it unconstrained} the components of
the canonical momentum
\be
     p_\mu = {{\p L}\oo {\p {\dot x}^\mu}} = {{\kappa {\dot x}^\mu}\oo 
    {{\dot x}^\mu {\dot x}_\mu- 1}}
\lbl{4.21}
\ee
are all independent. From the equations of motion it follows that
$p_\mu$ is constant of motion and so is its square
\be
     p^\mu p_\mu  = {{\kappa^2 {\dot x}^\mu {\dot x}_\mu}\oo
     \sqrt{{\dot x}^\nu {\dot x}_\nu - 1}} = M^2
\lbl{4.22}
\ee
Here $M^2$ is {\it not} a fixed constant entering the action, but it is
an arbitrary constant of motion. 

The theory based on the unconstrained action (\ref{4.20}) is equivalent to
the Stueckelberg theory based on the unconstrained action (\ref{4.15}).
Whilst in the former theory gauge is chosen according to eq.(\ref{4.14}),
in the latter theory gauge is chosen according to ${\dot s} = 1$.
Both theories are derived from the polyvector
action (\ref{4.1}), which---when eq.(\ref{4.4}) holds---is equivalent
to (\ref{4.7}). Although 
(\ref{4.7}) looks like an action which encompasses an extra dimension, 
this is not the case. In eq.(\ref{4.7}) $s$ {\it is not a variable due to
an extra dimension, it is a coordinate of $C$-space, that is the Clifford
manifold generated by a set of basis vectors spanning our spacetime.}

In $C$-space there is the {\it constraint}. Namely, the $C$-space momentum
$P_A = \p L/\p X^A$ is constrained to the ``mass shell""
\be
      P_A P^A = \kappa^2
\lbl{4.26}
\ee
Explicitly, for our particular case (\ref{4.4}) eq.(\ref{4.26}) reads as
eq.(\ref{4.19d}). That is, the four components $p_\mu$ and the extra component
$p_{(s)}$ (due to the extra variable $s$) altogether are constrained. But the
4-momentum $p_\mu$ alone, i.e., the momentum belonging to the reduced
action (\ref{4.15}) or (\ref{4.20}), is not constrained. And since
(\ref{4.15}) or (\ref{4.20}) describe a free particle (without interaction),
momentum $p_\mu$ is an arbitrary constant of motion.

{\it The $C$-space action} (\ref{4.1}) or (\ref{4.7}) is invariant under
arbitrary reparametrizations of the parameter $\tau$, and a consequence is the
existence of the constraint. This is analogous to the situation in the
ordinary theory of relativity. {\it Therefore in $C$-space we have
a ``block universe": everything is frozen once for all.}

{\it The reduced actions} (\ref{4.15}) and (\ref{4.20}) are not invariant under
reparametrizations of $\tau$, there is no constraint, and all four coordinates
$x^\mu$ evolve independently along $\tau$ which is a true evolution
parameter, identified in the case of eq.(\ref{4.15}) with the length of
the worldline in $C$-space or in the case of eq.(\ref{4.20})
with the extra variable $s$. So we have evolution
in 4-dimensional spacetime which is a section through $C$-space.

The enigma of why we feel the passage of time---a concept that does not
exist in the theory of relativity---is thus resolved, at least formally
(see also refs.\ci{Pavsic1,Pavsic2}), by postulating that
spacetime slice $M_4$ moves through $C$-space. A consequence is a genuine
dynamics ({\it relativistic dynamic}) on $M_4$. On the other hand,
{\it all the elegance of relativity is preserved, not in spacetime,
but in $C$-space.}

\section{Conclusion}

We have shifted the theory of relativity from the 4-dimensional spacetime to the
Clifford space. The latter space is a very natural---in fact, 
unavoidable---generalization of spacetime $V_4$. The geometry of spacetime is
described in terms of vectors which can be elegantly represented as
Clifford numbers. Once we have a set of four Clifford numbers $\gamma_\mu$
representing four independent basis vectors of $V_4$ we can automatically
generate a larger structure, namely the Clifford algebra ${\cal C}_4$ of
$V_4$ which encompasses multivectors. Since ${\cal C}_4$ is a manifold,
we call it the Clifford space or $C$-space. Owing to the fact, explored
in this paper, that the extended  object can be described by multivectors,
it is very natural to assume that  physics fundamentally takes
place in $C$-space. This has many far reaching physical consequences,
some of which are described in this paper (for the rest see also 
ref.\ci{Pavsic1}).

Relativity in $C$-space treats the multivector coordinates of an extended
object as the  {\it degrees of freedom} and predict that the
associated multivector (holographic) velocities can be different from zero.
In particular, when the bivector and higher multivector velocities are zero,
then the new theory is
indistinguishable from the existing theory of relativity. All the
predictions of the ordinary theory of relativity are then preserved in the new
theory. However, from the point of view of the relativity in $C$-space,
the ordinary relativity has restricted validity: it holds only when
${\dot X}^{\mu \nu}$, ${\dot X}^{\mu \nu \alpha}$, etc., are zero.
When they are different from zero we have
violation of the ordinary relativity and Lorentz invariance (in $M_4$)
in the sense that the worldlines of freely moving physical objetcs are no longer
confined to $M_4$, but they escape into the larger space, namely $C$-space.
The transformations that relate such worldlines are in general not the
usual Lorentz transformations, but the generalized Lorentz transformations,
namely the rotations in $C$-space (\ref{24A}). Lorentz group is thus no longer
the exact symmetry group of physical objetcs and speed of light not the
maximum speed. 
The possibility that Lorentz invariance might be violated has been
widely discussed during last years \ci{Lorentz} in various contexts,
especially within noncommutative geometries \ci{Noncommute}.

Relativity in $C$-space implies that in Minkowski space $M_4$ there is
the genuine dynamics---``relativistic dynamics"---as advocated by
Stueckelberg and his followers \ci{Stueckelberg}. The Stueckelberg theory,
since being unconstrained, has many desirable properties, especially when
quantized  \ci{Stueckelberg,Pavsic3} and provides a natural explanation 
\ci{Pavsic1,Pavsic2,Pavsic4} of 
``the passage of time", whilst the more general theory, namely the relativity
in $C$-space retains all the nice features and elegance of the theory of 
relativity, including reparametrization invariance, ``block universe", etc..
The objects, when viewed from the $C$-space perspective are infinitely
extended, frozen, ``world lines" $X^A (\tau) \equiv X^{\mu_1 ...\mu_r}(\tau)$,
analogous to the world lines of the ordinary relativity. When viewed from the
perspective of 4-dimensional spacetime (which is just a subspace of the 
``full" space, namely $C$-space) the objects have {\it finite} extensions
both in space-like and time-like direction. They move in spacetime, and during
the motion their extensions, including the time-like extensions, change
with $\tau$. From the point of view of the new theory, $p$-branes (including
strings)---whose ($p+1$)-dimensional worldsheets are infinitely extended 
objects along time-like directions---
are merely idealized objects to which the real physical objects approach
when the evolution parameter $\tau$ goes
to infinity. At finite $\tau$, all objects in spacetime are
predicted to have finite extension. 

All this is just the start. A Pandora box of fascinating new possibilities
is open, some of them are touched in a recent book \ci{Pavsic1} and refs. 
\ci{Castro-Pavsic}. A very promising perspective is in exploiting the well 
known fact \ci{Ideals} that spinors are members of left and right minimal ideals
of Clifford algebra. This means that spinors are nothing but special
kind of polyvectors. Therefore the presence of spinors is automatically
included in the coordinate polyvector $X=X^A E_A$. We therefore expect that
suitable generalizations of the action (\ref{25}) (see \ci{Pavsic1}) will
provide a description of spinning $p$-branes and super $p$-branes
including superstrings, superparticles or spinning strings and spinning
particles. It seems very likely that further development of the theory
based on $C$-space will lead us towards M-theory 
and towards the unified theory of the known fundamental interactions.


\begin{thebibliography}{12}

\bi{Hestenes} D. Hestenes, {\it Space-Time Algebra} (Gordon and Breach,
New York, 1966); D. Hestenes {\it Clifford Algebra to Geometric Calculus}
(D. Reidel, Dordrecht, 1984)\bi{Pezzaglia} W.M. Pezzaglia, {\it A Clifford 
Algebra Multivector
Reformulation of Field Theory}, Dissertation (University of California,
Davis, 1983); W. M. Pezzaglia Jr, {\it Classification of 
Multivector Theories and
Modification o f the Postulates of Physics} [arXiv: gr-qc/9306006];
{\it Polydimensional Relativity, a Classical 
Generalization of the
Automorphism Invariance Principle}, [arXiv:gr-qc/9608052];
{\it Physical Applications of a Generalized Clifford
Calculus: Papapetrou Equations and Metamorphic Curvature},
[arXiv:gr-qc/9710027];  
W. M. Pezzaglia Jr nad J. J. Adams, {\it Should Metric Signature Matter 
in Clifford Algebra Formulation of
Physical Theories?}, [arXiv:gr-qc/9704048];\\   
W. M. Pezzaglia Jr and A. W. Differ, {\it A Clifford Dyadic Superfield from
Bilateral Interactions of Geometric Multispin Dirac Theory},
[arXiv:gr-qc/9311015];
W. M. Pezzaglia Jr, {\it Dimensionally Democratic Calculus and
Principles of Polydimensional Physics}, [arXiv:gr-qc/9912025];

\bi{Castro} C. Castro, {\it Chaos, Solitons and Fractals} {\bf 11}, 1721 (2000)
[arXiv:hep-th/9912113];
{\it J. Chaos, Solitons and Fractals}, {\bf 12}, 1585 (2001)
[arXiv:physics/0011040]

\bi{Pavsic1} 
M. Pav\v si\v c,
%``Clifford Algebra Based Polydimensional Relativity and Relativistic Dynamics,''
{\it Found.\ Phys.}\  {\bf 31}, 1185 (2001) [arXiv:hep-th/0011216];
%%CITATION = HEP-TH 0011216;%%;
M. Pav\v si\v c, {\it The Landscape of Theoretical Physics:
A Global View. From Point Particles to the Brane World and Beyond,
in Search of a Unifying Principle} (Kluwer, Dordrecht, 2001)

\bi{Stueckelberg} V. Fock, {\it Phys. Z. Sowj.} {\bf 12}, 404 (1937);
E.C.G. Stueckelberg, {\it Helv. Phys. Acta}, 
{\bf 14}, 322 (1941); {\bf 14}, 588 (1941); {\bf 15}, 23 (1942);
R. P. Feynman {\it Phys. Rev}, {\bf 84}, 108 (1951)
\item L. P. Horwitz and C. Piron, {\it Helv. Phys. Acta}, {\bf 46}, 316 (1973);
L. P. Horwitz and F. Rohrlich, {\it Physical Review D} {\bf 24}, 1528 (1981);
{\bf 26}, 3452 (1982); L. P. Horwitz, R. I. Arshansky and A. C. Elitzur
{\it Found. Phys} {\bf 18}, 1159 (1988); R. Arshansky, L. P. Horwitz and
Y. Lavie, {\it Foundations of Physics} {\bf 13}, 1167 (1983);
L. P. Horwitz, in {\it Old and New Questions in Physics, Cosmology,
Philosophy and Theoretical Biology} (Editor Alwyn van der Merwe, Plenum,
New York, 1983); L. P. Horwitz and Y. Lavie, {\it Physical Review D} {\bf 26},
819 (1982);
L. Burakovsky, L. P. Horwitz and W. C. Schieve, {\it Physical Review D}
{\bf 54}, 4029 (1996); L. P. Horwitz and W. C. Schieve,
{\it Annals of Physics} {\bf 137}, 306 (1981);
J.R.Fanchi, {\it Phys. Rev. D} {\bf 20}, 3108 (1979);
see also the review J.R.Fanchi, {\it Found. Phys.} {\bf 23},
287 (1993), and many references therein; J. R. Fanchi {\it Parametrized
Relativistic Quantum Theory} (Kluwer, Dordrecht, 1993);
M. Pav\v si\v c, {\it Found. Phys.} {\bf 21}, 1005 (1991);
M. Pav\v si\v c,{\it Nuovo Cim.} {\bf A104}, 1337 (1991)

%\cite{Pavsic:1976ta}
\bibitem{Pavsic:1976ta}
M. Pav\v si\v c,
%``Unified Theory Of Gravitation And Electromagnetism Based On 
%The Conformal Group SO(4,2),''
{\it Nuovo Cim.\ B} {\bf 41}, 397 (1977);
%%CITATION = NUCIA,B41,397;%%
%\cite{Pavsic:1980gy}
%\bibitem{Pavsic:1980gy}
M.~Pav\v si\v c,
%``Introducing The Dilatational Degree Of Freedom: Special Relativity In V(6),''
{\it J.\ Phys.\ A} {\bf 13}, 1367 (1980).
%%CITATION = JPAGB,A13,1367;%%

\bi{Aurilia} A. Aurilia, C. Castro and M. Pav\v si\v c, in preparation.

\bi{alpha} J.K. Webb, V.V. Flambaum, C.W.Churchill, M.J. Drinkwater, and
J.D. Barrow, {\it Phys. Rev. Lett.} {\bf 82}, 884 (1999);
J.K. Webb, M.T. Murphy, V.V. Flambaum, J.D. Dzuba, J.D. Barrow, C.W. Churchill,
J.X. Prochaska, and A.M. Wolfe, {\it Phys. Rev. Lett.}
{\bf 87}, 091391 (2001); M.T. Murphy, J.K. Webb, V.V. Flambaum, V.A.
Dzuba, C.W. Churchill, J.X. Prochaska, J.D. Barrow, and A.M. Wolfe,
{\it Mon. Not. R. Astron. Soc.} {\bf 327}, 1208 (2001)

\bi{Pavsic2}
%\cite{Pavsic:1995kc}
%\bibitem{Pavsic:1995kc}
M. Pav\v si\v c,
%``On the resolution of time problem in quantum gravity 
%induced from unconstrained membranes,''
{\it Found.\ Phys.}\  {\bf 26}, 159 (1996)
[arXiv:gr-qc/9506057].
%%CITATION = GR-QC 9506057;%%

\bi{Carlini} J.Greensite, {\it Class. Quant. Grav.} {\bf 13}, 1339 (1996);
{\it Phys. Rev. D} {\bf 49}, 930 (1994); A. Carlini and J. Greensite,
{\it Phys. Rev. D} {\bf 52}, 936 (1995); {\bf 52}, 6947 (1995); {\bf 55},
3514 (1997)

\bi{Howe-Tucker}P.S. Howe and R.W. Tucker, {\it J. Phys. A: Math. Gen.} {\bf 10},
L155 (1977); A. Sugamoto, {\it Nuclear Physics B} {\bf 215}, 381 (1983);
E. Bergshoeff, E. Sezgin and
P.K. Townsend, {\it Physics Letter B} {\bf 189}, 75 (1987);
A. Achucarro, J.M. Evans, P.K. Townsend and D.L. Wiltshire, {\it
Physics Letters B} {\bf 198}, 441 (1987);
%\cite{Pavsic:ny}
%\bibitem{Pavsic:ny}
M. Pav\v si\v c,
%``Generalization Of The Bdhp String Action To Membranes Of Any Dimension 
%In Curved Space-Time,''
{\it Class.\ Quant.\ Grav.}\  {\bf 5}, 247 (1988);
%%CITATION = CQGRD,5,247;%%
%\cite{Pavsic:ei}
%\bibitem{Pavsic:ei}
M. Pav\v si\v c,
%``Phase Space Action For Minimal Surfaces Of Any Dimension In Curved 
%Space-Time,''
{\it Phys.\ Lett.\ B} {\bf 197}, 327 (1987).
%%CITATION = PHLTA,B197,327;%%

\bi{Lorentz} D. Colladay and P.McDonald, {\it Journ. Math. Phys.} {\bf 43},
3554 (2002), and references therein

\bi{Noncommute} A. Connes, ``C*-Alg\` ebres et G\' eometrie Diff\' erentielle",
{\it C.R. Acad. Sci. Paris} {\bf 290}, 599 (1980); 
For a recent review see e.g., J. Madore,{\it  An Introduction to Noncommutative 
Differential Geometry and Physical Applications}, (Cambridge University Press,
2000)

\bi{Pavsic3}
%\cite{Pavsic:1995gc}
%\bibitem{Pavsic:1995gc}
M. Pav\v si\v c,
%``The Classical and quantum theory of relativistic p-branes without 
%constraints,''
{\it Nuovo Cim.\ A} {\bf 108}, 221 (1995)
[arXiv:gr-qc/9501036];
%%CITATION = GR-QC 9501036;%% 
%\cite{Pavsic:je}
%\bibitem{Pavsic:je}
M. Pav\v si\v c,
%``Relativistic P-Branes Without Constraints And Their Relation To 
%The Wiggly Extended Objects,''
{\it Found.\ Phys.}\  {\bf 25}, 819 (1995);
%%CITATION = FNDPA,25,819;%%
%\cite{Pavsic:1997eu}
%\bibitem{Pavsic:1997eu}
M. Pav\v si\v c,
%``The Dirac-Nambu-Goto p-branes as particular solutions to a 
%generalized,  unconstrained theory,''
{\it Nuovo Cim.\ A} {\bf 110}, 369 (1997)
[arXiv:hep-th/9704154];
%%CITATION = HEP-TH 9704154;%%
%\cite{Pavsic:yn}
%\bibitem{Pavsic:yn}
M. Pav\v si\v c,
%``Formulation Of A Relativistic Theory Without Constraints,''
{\it Found.\ Phys.}\  {\bf 28}, 1443 (1998);
%%CITATION = FNDPA,28,1443;%%
%\cite{Pavsic:yp}
%\bibitem{Pavsic:yp}
M. Pav\v si\v c,
%``Parametrized Field Theory,''
{\it Found.\ Phys.}\  {\bf 28}, 1453(1998) .
%%CITATION = FNDPA,28,1453;%%

\bi{Pavsic4} 
%\cite{Pavsic:1990th}
%\bibitem{Pavsic:1990th}
M. Pav\v si\v c,
%``On The Interpretation Of The Relativistic Quantum Mechanics 
%With Invariant Evolution Parameter,''
{\it Found.\ Phys.}\  {\bf 21}, 1005 (1991).
%%CITATION = FNDPA,21,1005;%%

\bi{Castro-Pavsic}
%\cite{Castro:2001gf}
%\bibitem{Castro:2001gf}
C.~Castro and M.~Pav\v si\v c,
%``Higher Derivative Gravity and Torsion from the Geometry of C-spaces,''
{\it Phys.\ Lett.\ B} {\bf 539}, 133 (2002)
[arXiv:hep-th/0110079];
%%CITATION = HEP-TH 0110079;%%
%\cite{Castro:2002kt}
%\bibitem{Castro:2002kt}
C.~Castro and M. Pav\v si\v c,
{\it Clifford Algebra of Spacetime and the conformal Group,}
[arXiv:hep-th/0203194].
%%CITATION = HEP-TH 0203194;%%

\bi{Ideals} S. Teitler, {\it Supplemento al Nuovo Cimento} {\bf III}, 1 (1965);
{\it Supplemento al Nuovo Cimento} {\bf III}, 15 (1965);
{\it Journal of Mathematical Physics} {\bf 7}, 1730 (1966);
{\it Journal of Mathematical Physics} {\bf 7}, 1739 (1966);
L. P. Horwitz, {\it J. Math. Phys.} {\bf 20}, 269 (1979);
H. H. Goldstine and L. P. Horwitz, {\it Mathematische Annalen} {\bf 164},
291 (1966)


\end{thebibliography}
\end{document}